\documentclass[aps,prl,reprint,superscriptaddress,10pt]{revtex4-1}
\usepackage{amsmath}
\usepackage{hyperref}
\usepackage{graphicx}
\usepackage{xcolor}
\usepackage{epstopdf}
\usepackage{verbatim}
\usepackage{physics}
\usepackage{pgfplots}
\usepackage[normalem]{ulem}

\usepackage{tikz}
\usetikzlibrary{shapes.geometric, arrows}
\tikzstyle{process} = [rectangle, minimum width=3cm, minimum height=1cm, text centered, draw=white,fill=white]
\tikzstyle{arrow} = [thick,->,>=stealth]

\begin{document}

\title{Single-shot measurements of phonon number states using the Autler-Townes effect}

\author{Marion~Mallweger}
\email[]{marion.mallweger@fysik.su.se}
\affiliation{Department of Physics, Stockholm University, SE-106 91 Stockholm,
Sweden}
\author{Murilo Henrique de Oliveira}
\affiliation{Departamento de F\'isica, Universidade Federal de S\~ao Carlos, 13565-905 S\~ao Carlos, SP, Brazil}
\author{Robin~Thomm}
\affiliation{Department of Physics, Stockholm University, SE-106 91 Stockholm,
Sweden}
\author{Harry~Parke}
\affiliation{Department of Physics, Stockholm University, SE-106 91 Stockholm,
Sweden}
\author{Natalia~Kuk}
\affiliation{Department of Physics, Stockholm University, SE-106 91 Stockholm,
Sweden}
\author{Gerard~Higgins}
\affiliation{Department of Physics, Stockholm University, SE-106 91 Stockholm,
Sweden}
\affiliation{Department of Microtechnology and Nanoscience (MC2), Chalmers University of Technology, SE-4DS 96 Gothenburg, Sweden}
\author{Romain Bachelard}
\affiliation{Departamento de F\'isica, Universidade Federal de S\~ao Carlos, 13565-905 S\~ao Carlos, SP, Brazil}
\affiliation{Universit\'e C\^ote d'Azur, CNRS, Institut de Physique de Nice, 06560 Valbonne, France}
\author{Celso Jorge Villas-Boas}
\affiliation{Departamento de F\'isica, Universidade Federal de S\~ao Carlos, 13565-905 S\~ao Carlos, SP, Brazil}
\author{Markus~Hennrich}
\affiliation{Department of Physics, Stockholm University, SE-106 91 Stockholm,
Sweden}

\begin{abstract}
We present a single-shot method to measure motional states in the number basis. The technique can be applied to systems with at least three non-degenerate energy levels which can be coupled to a linear quantum harmonic oscillator, such as in trapped ion experiments. The method relies on probing an Autler-Townes splitting that arises when two levels are strongly coupled via a phonon-number changing transition. We demonstrate the method using a single trapped ion and show that it may be used in a non-demolition fashion to prepare phonon number states. We also show how the Autler-Townes splitting can be used to measure phonon number distributions. 
\end{abstract}

\maketitle

The common approach to trapped ion quantum information processing is to use electronic states to store information while the motional modes shared by a chain of ions enable entangling operations~\cite{Leibfried2003}. However, the motional modes can play a more active role. For instance, the motional degree of freedom can be used for storing quantum information~\cite{Fluhmann2019}, allowing for continuous variable quantum information processing with trapped ions. The motional modes are also a very important tool in quantum logic spectroscopy~\cite{Schmidt2005}, which enables the implementation of precise atomic clocks~\cite{Brewer2019}. Also, in metrology an advantage can be achieved via non-classical states of ion motion \cite{McCormick2019, Wolf2019, Zhang2018}.
On the more fundamental side, trapped ion motion acts as the working medium in studies of quantum thermodynamics~\cite{An2014, Rossnagel2016, vonLindenfels2019}.
Investigations of the dynamics of phonon pair creation upon trap potential changes are a possibility to simulate particle creation, thus creating a link between quantum information processing and cosmology~\cite{Wittemer2019}. Finally, the measurements of local phonons and their tracking enables quantum simulations in motional degrees of freedom~\cite{Ohira2019, Tamura2020}.

Trapped ion motion can be measured in various ways~\cite{Meekhof1996, Leibfried1996, Shen2014, An2014, Um2016, Ding2017, Ohira2019, Meir2017}, including via cross-Kerr nonlinearity~\cite{Ding2017, Roos2008, Marquet2003} and composite pulse sequences~\cite{Ohira2019}. There are also schemes using rapid adiabatic passage (RAP)~\cite{poschinger2012,Watanabe2011} and stimulated Raman adiabatic passage (STIRAP)~\cite{Gebert2016} sequences or poly-chromatic amplitude modulated beams~\cite{Muller2015}. However, most detection methods destroy the state throughout the measurement process.

Schemes for resolving Fock states in a non-demolition manner have been proposed and measured using the dispersive AC Stark shift on Rydberg atoms interacting with photons in a cavity~\cite{Brune1990,Guerlin2007}. Here, the measurement relies on a Ramsey type experiment to measure the phase shift introduced by the cavity photons. The same dispersive interaction has been used in superconducting circuits~\cite{Schuster2007,Arrangoiz2019}, where a qubit is coupled to the mode of a microwave cavity in a strong dispersive interaction regime. The AC Stark shift resulting from the coupling splits the qubit spectrum, turning it into an anharmonic ladder where the dressed states energies are proportional to the dispersive coupling rate and, consequently, to the number of photons in the system.

In this work, we introduce a novel technique that is based on the Autler-Townes effect~\cite{Autler1955} and can measure in a single shot a motional mode in the number (Fock) basis.
Our approach also allows for quantum non-demolition measurement of motional Fock states and can be used to determine the phonon distribution. We demonstrate this method using a single $\mathrm{^{88}Sr^+}$ ion trapped in a linear Paul trap.

\hfill

The Autler-Townes effect is commonly probed in three-level systems and in this work we denote these levels as $\ket{S}$, $\ket{D}$, and $\ket{D'}$.
The coupling of the two levels, $\ket{S}$ and $\ket{D'}$, via a laser field is described by the following Hamiltonian
\begin{equation}
H = \frac{\hbar}{2} \begin{pmatrix}
0 & \Omega_C\\
\Omega_C & 2\Delta_C
\end{pmatrix},
\end{equation}
where $\Omega_C$ is the coupling strength and $\Delta_C=\omega_L-\omega_0$ is the detuning of the laser field frequency $\omega_L$ from the transition resonance $\omega_0$.
When the laser field is resonant ($\Delta_C=0$) the dressed eigenstates with eigenenergies $\pm  \hbar \Omega_C/2$ are $\left(\ket{S} \pm \ket{D'}\right)/\sqrt{2}$, respectively.

These two eigenstates can be probed from the third level $\ket{D}$. Namely, if level $\ket{S}$ is strongly coupled to the level $\ket{D'}$ the spectral resonance for the transition from level $\ket{D}$ to level $\ket{S}$ is split into a doublet.
By driving the system resonantly to one of the doublet peaks at $\pm  \hbar \Omega_C/2$ it is possible to excite the respective dressed eigenstate $\left(\ket{S} \pm \ket{D'}\right)/\sqrt{2}$.
This is called the Autler-Townes effect \cite{Autler1955}.
\begin{figure}[t]
\centering
\includegraphics[width=\columnwidth]{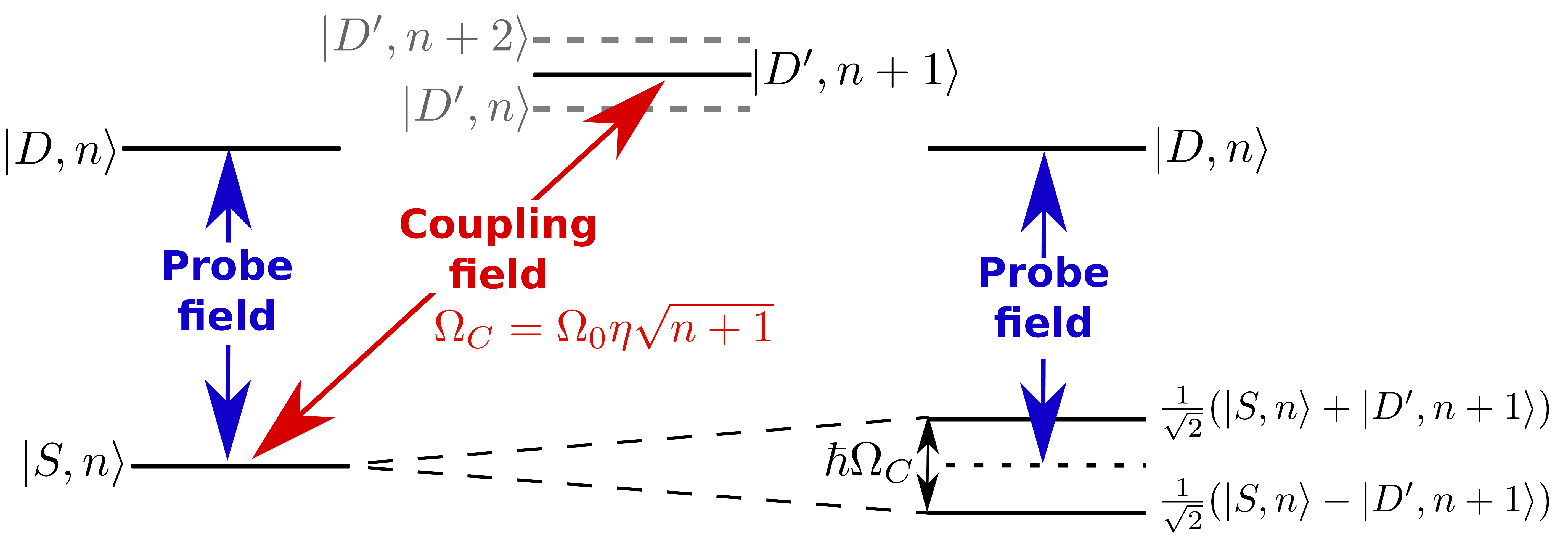}
\caption[width=\textwidth]{Autler-Townes scheme for phonon number detection. (a) The $\ket{D,n} \leftrightarrow \ket{S,n}$ resonance is weakly probed while levels $\ket{S,n}$ and $\ket{D',n+1}$ are strongly coupled via a laser on a blue sideband (BSB) transition. (b) The system described in terms of dressed states. The splitting between the dressed states reveals the $\ket{S,n} \leftrightarrow \ket{D',n+1}$ coupling strength $\Omega_C$. The splitting behaviour is similar when coupling to a red sideband (RSB) transition.
}
\label{fig_ATS_schematic}
\end{figure}

Here, we show that if the internal levels are coupled to a motional degree of freedom the Autler-Townes effect can be used to probe motional number states.
When $\ket{S,n} \leftrightarrow \ket{D',n\pm 1}$ is a phonon-number-changing transition, with $n$ the respective phonon number states, $\Omega_C$ becomes sensitive to the population of the motional mode \cite{sorensen1999}, as it is depicted in Fig.~\ref{fig_ATS_schematic}. The splitting between the Autler-Townes doublet peaks is proportional to the strength of the $\ket{S,n} \leftrightarrow \ket{D',n \pm 1}$ coupling strength, $\Omega_C(n)$.
This allows us to use the Autler-Townes doublet to probe the phonon number. In Fig.~\ref{fig_ATS_schematic} the Autler-Townes effect is plotted for the coupling to the blue sideband transition (BSB) of a motional mode, where the couplings scales with $\sqrt{n+1}$. An analogue scheme can be used for a red sideband transition (RSB) with a scaling of $\sqrt{n}$.
\begin{figure}
\centering
    \includegraphics[width=\columnwidth]{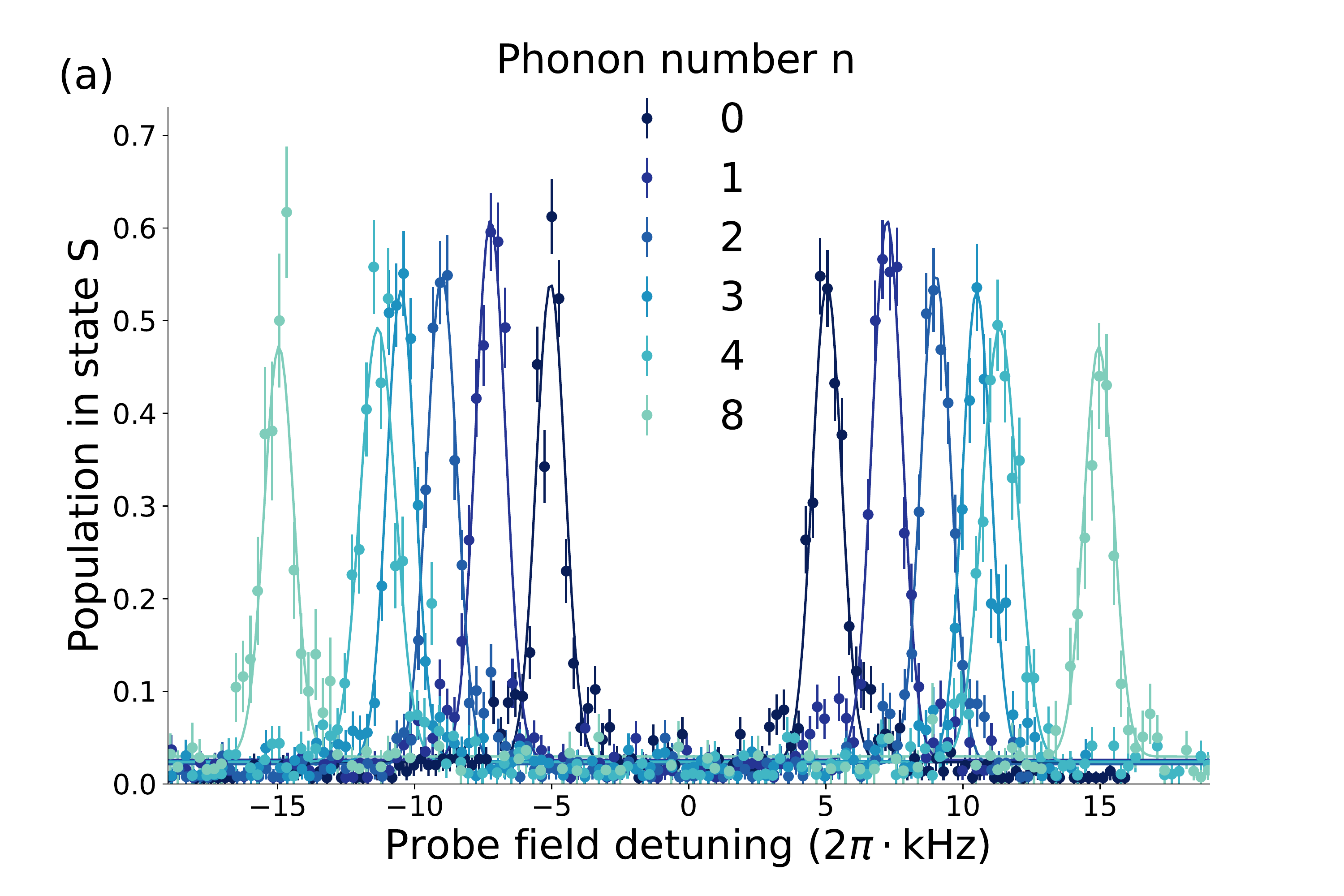}  
    \label{fig_data_BSB}
    
    \includegraphics[width=\columnwidth]{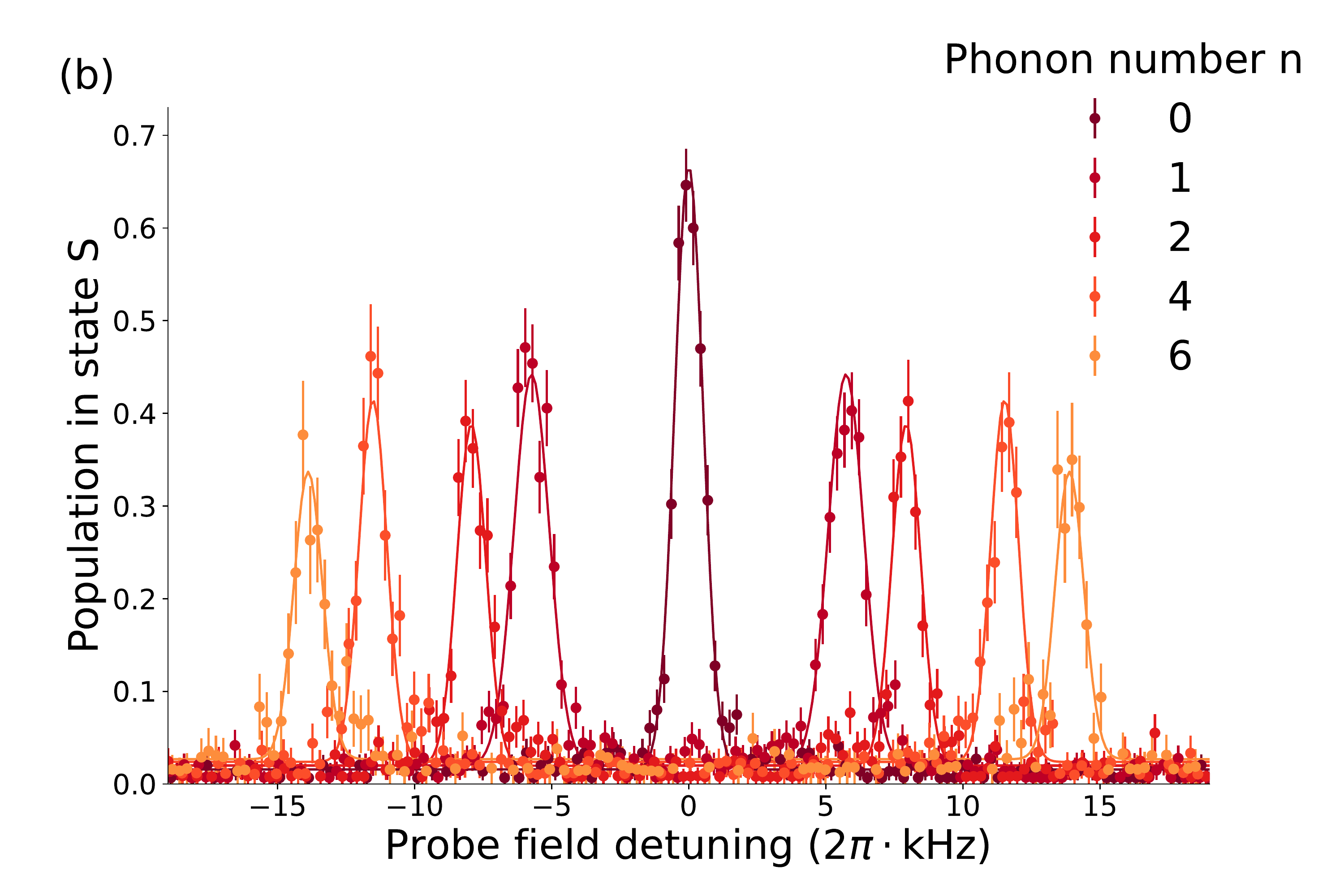}  
    \label{fig_data_RSB}
    
    \includegraphics[width=\columnwidth]{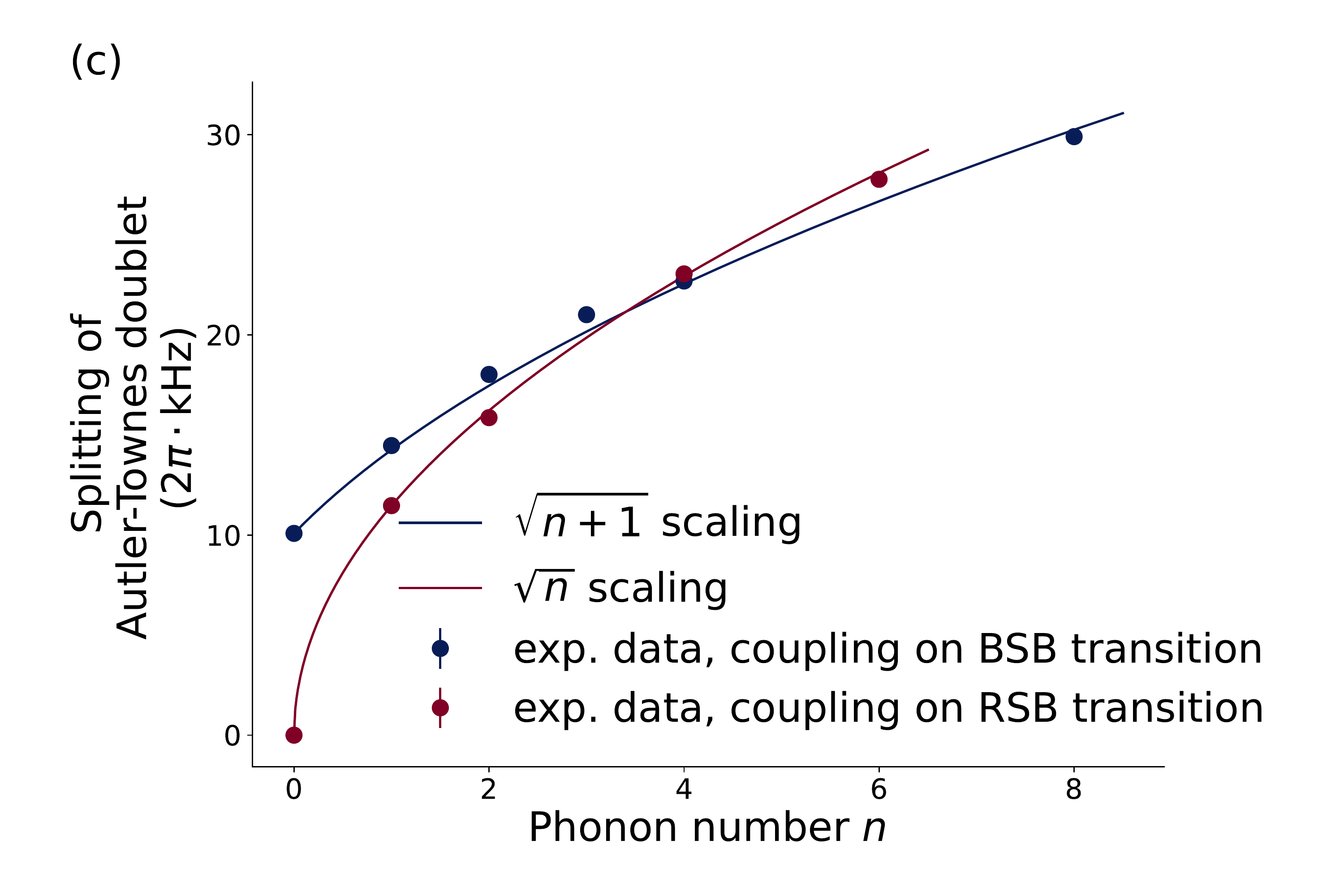}
    
\caption[width=\textwidth]{Splitting of the Autler-Townes doublet. When the coupling field is resonant to a phonon-number-changing transition the splitting of the Autler-Townes doublet depends on the number of phonons in the system.
(a)~For the coupling field resonant to a BSB transition the splitting scales with the phonon number $n$ as $\sqrt{n+1}$. (b)~For the coupling field resonant to a RSB transition the splitting scales with $\sqrt{n}$. Error bars represent quantum projection noise (68\% confidence intervals). The lines are fits to the data with the amplitude and the splitting as a fitting parameters. (c)~Splitting of the Autler-Townes doublet extracted from the fitting curves in (a) and (b). The scalings of the splitting are described by $\sqrt{n+1}$ (BSB transition) and $\sqrt{n}$ (RSB transition), respectively.}
\label{fig_AT_results}
\end{figure}

We demonstrated this phonon number-dependent Autler Townes effect by preparing a single trapped $\mathrm{^{88}Sr^+}$ ion in electronic state $\ket{D,n}$, where $n$ describes the number state of one of the radial modes. For the measurements presented we employed states $\ket{D} \equiv 4{}^2D_{5/2} \:, m_J=-\frac{1}{2}$, $\ket{S} \equiv 5{}^2S_{1/2} \:, m_J=-\frac{1}{2}$ and $\ket{D'} \equiv 4{}^2D_{5/2} \:, m_J=-\frac{3}{2}$. We probed the $\ket{D,n} \leftrightarrow \ket{S,n}$ carrier resonance while strongly coupling the BSB transition, $\ket{S,n}\leftrightarrow\ket{D',n+1}$.
The resultant spectrum displays an Autler-Townes doublet with a phonon-number dependent splitting, see Fig.~\ref{fig_AT_results}(a).

When applying the strong coupling field to a RSB transition $\ket{S,n}$ to $\ket{D',n-1}$, the resultant spectrum also displays an Autler-Townes doublet however with a different coupling dependent scaling than the BSB transition, see Fig.~\ref{fig_AT_results}(b). For both experiments the peak height of the Autler-Townes doublet decreases with an increasing phonon number because the coupling strength of the probe field changes proportional to $(1-\eta^2n)$.

We extract the doublet splittings by fitting the data presented in Figs.~\ref{fig_AT_results}(a) and (b). The fit confirms the respective $\sqrt{n+1}$ and $\sqrt{n}$ scalings, as shown in Fig.~\ref{fig_AT_results}(c).
The scalings of the doublet splittings are consistent with the strengths of the phonon-number changing transitions for 
laser-ion interaction in the Lamb-Dicke regime \cite{Leibfried2003}.  For both Jaynes-Cummings type interactions, the coupling to the RSB and the BSB transition, the scaling of the splitting is proportional to the initial coupling strength $\Omega_0$. In Fig.~\ref{fig_AT_results}, the splitting of the RSB takes larger values for increasing phonon number because of the  higher value of $\Omega_0$ for this scan.

\begin{figure}[ht]
    \centering
    \includegraphics[width=\columnwidth]{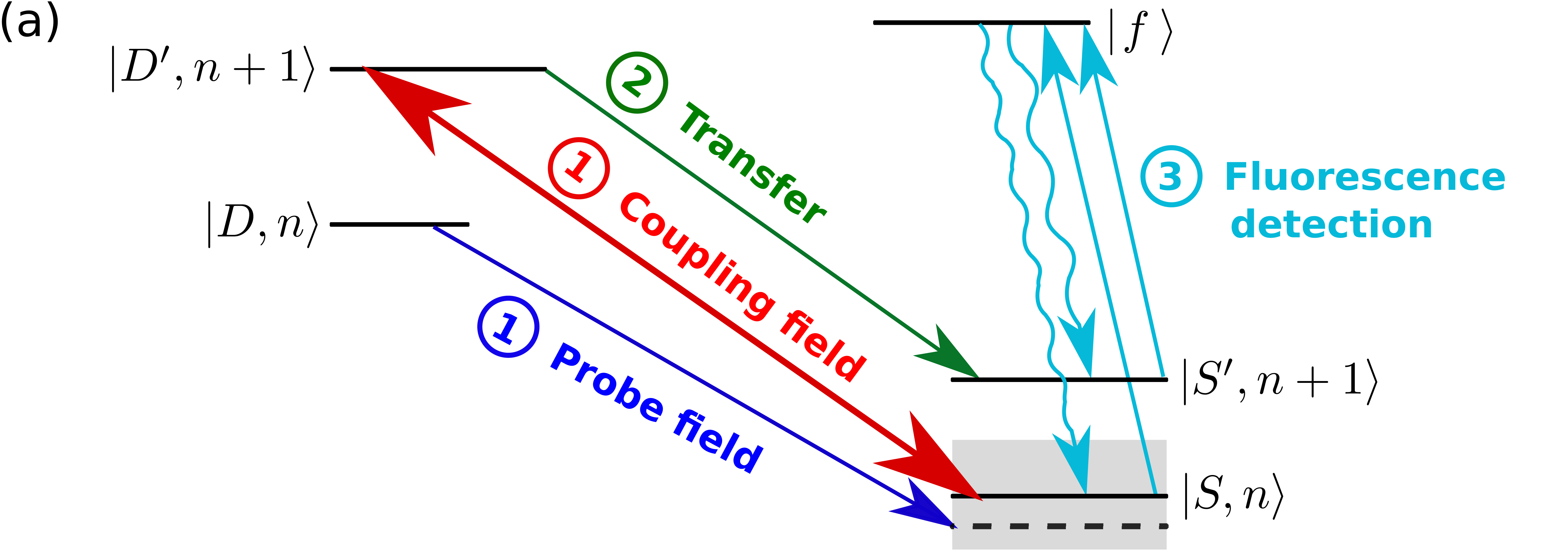}
    \label{fig_sequence}
    
    \includegraphics[width=\columnwidth]{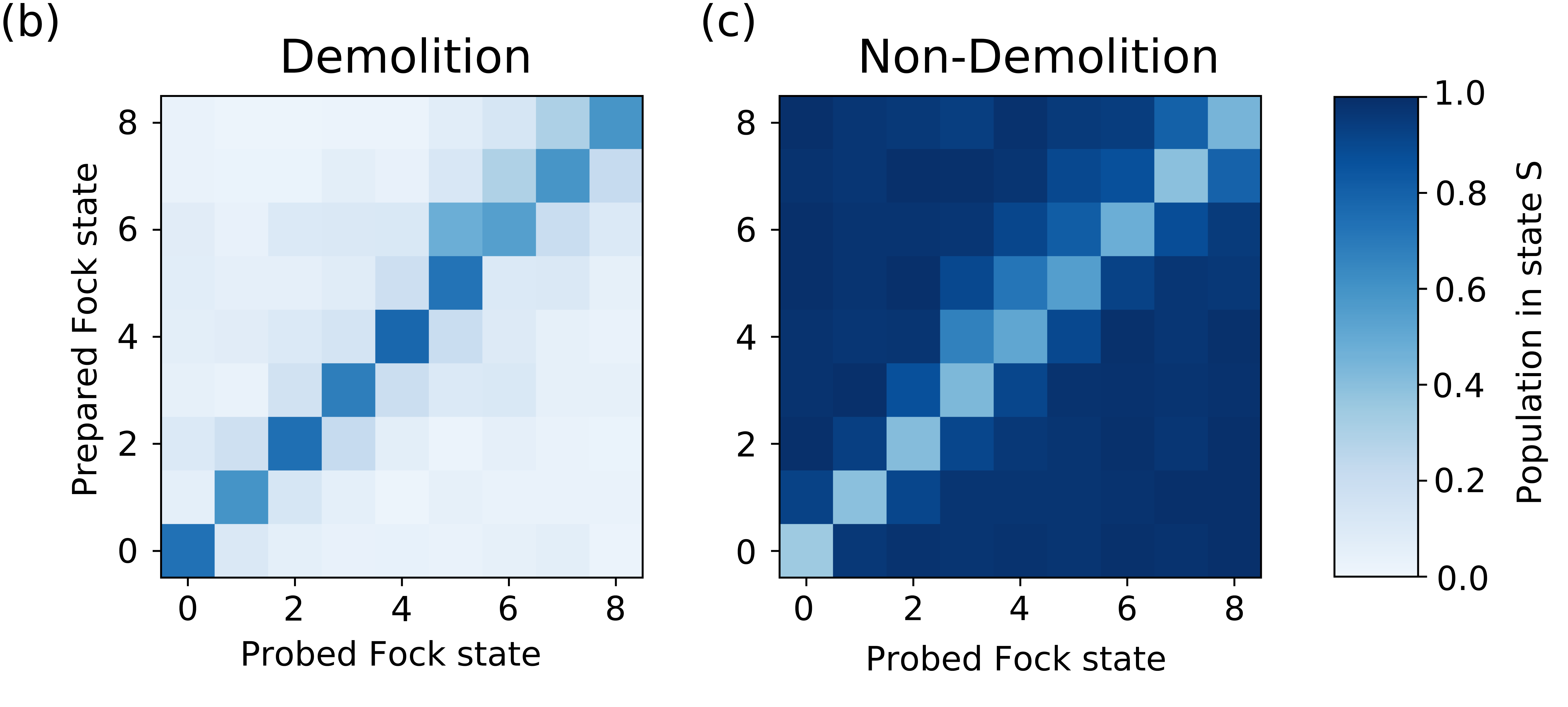}
    
\caption[width=\textwidth]{(a) Measurement sequence: The ion is initialised in $\ket{D,n}$. In the first step, a $\pi$ pulse is applied on the $\ket{D} \leftrightarrow \ket{S}$ carrier transition while a phonon-number-changing transition is strongly coupled (in the figure a BSB transition is shown). During this step the probe field is detuned such that the $\ket{D} \rightarrow \ket{S}$ transfer only occurs if the prepared Fock state $\ket{n}$ equals the probed Fock state $\ket{m}$.
In the second step, any population in $\ket{D'}$ is transferred to $\ket{S'}$ before both $S$ states are coupled to the fluorescing state $\ket{f}$. Finally, detection of fluorescence indicates $n=m$.
(b)~Experimental demonstration of the sequence in (a). The ion was prepared in different Fock states, and when the probed Fock state matched the prepared one, fluorescence was detected.
(c)~Experimental results when the transfer step is applied to the $|D\rangle \leftrightarrow |S\rangle$ carrier transition, enabling a non-destructive measurement of the ion motion in the Fock basis.
}
\label{fig_fock_detect}
\end{figure}

The phonon-state dependent Autler-Townes splitting can be used to detect the occupation of a particular phonon number state with almost unit efficiency. The phonon detection pulse sequence is shown in Fig.~\ref{fig_fock_detect}(a). To prepare phonon number states we cooled the ion to its motional ground state and initialised the state $\ket{S,n=0}$ before iteratively adding phonons one by one. The phonon addition is performed by applying $\pi$ pulses first on the BSB transition to state $\ket{D, n+1}$ and then on the carrier transition back to state  $\ket{S, n+1}$, as described in \cite{Meekhof1996}. To achieve highly efficient preparation of phonon number states we included multiple state-dependent fluorescence detection steps. These were performed on the $\ket{S}$ to $\ket{f} \equiv 5{}^2P_{1/2}$ transition, during the phonon number preparation, as described in the supplemental material of ref.~\cite{Higgins2019}. Absence of fluorescence indicates successful addition of a phonon.

Before applying the phonon detection sequence, we used a $\pi$ pulse to initialize the ion in state $\ket{D,n}$. A $\pi$ pulse then drives the transition to one of the Autler-Townes doublet states for the phonon number state to be detected. Other phonon number states are off-resonant and will not be excited. At the end of the Autler-Townes phonon detection sequence the ion is in a superposition of the states $\ket{S,n}$ and $\ket{D',n+1}$, however, the fluorescence detection only shows a signal if the ion is in the $S$ manifold. Therefore, to increase the signal strength from the $S$ manifold at the end of the phonon detection sequence as a transfer step a $\pi$ pulse was applied on the transition from $\ket{D',n+1}$ to $\ket{S',n+1}$ (with $\ket{S'} \equiv 5{}^2S_{1/2} \:, m_J=\frac{1}{2}$).
Fluorescence detection was used to distinguish population in $|S\rangle$ and $\ket{S'}$ from population in $|D\rangle$ and $|D'\rangle$.  

Using this pulse scheme (Fig.~\ref{fig_fock_detect}(a)), the ions's motional state can be detected by probing discrete frequencies corresponding to the phonon-number-dependent peak position. We prepared and probed the ion in different phonon number states, ranging from 0 and 8, as shown in Fig.~\ref{fig_fock_detect}(b). If the probe field is not resonant to the probed phonon-dependent Autler-Townes transition, the ion remains undisturbed in $|D,n\rangle$. If the probe field drives the phonon-dependent Autler-Townes transition the final state is a superposition of the states $\ket{S,n}$ and $\ket{S',n+1}$. This outcome is detected by fluorescence. However, fluorescence detection involves the scattering of many photons and, therefore, changes the motional state, corresponding to a destructive (demolition) detection of the phonon state. The method may be repeated (without previous re-initialisation) to test for different phonon numbers sequentially until a positive result is achieved.
Each particular test relies on a priori knowledge of the expected peak positions. These can be determined from spectra as in Fig.~\ref{fig_AT_results} or by measuring the frequency of Rabi oscillations on the BSB transition for a ground-state-cooled ion.

In the method described above the nature of the detection scheme prevents any recovery of the previous state of the system once the correct phonon number state has been identified. By changing the transfer step (Fig.~\ref{fig_fock_detect}(a)) to a $\pi$-pulse on the $|S\rangle \leftrightarrow |D\rangle$ carrier transition, the absence of fluorescence indicates a positive result, as shown by the experimental data in Fig.~\ref{fig_fock_detect}(c).
This alternate method enacts a non-demolition measurement in the phonon-number basis. The final state of the system after detection is therefore a superposition of $\ket{D,n}$ and $\ket{D',n+1}$. From here the ion could be prepared in the measured phonon number state using the scheme presented in the supplementary material.

\hfill

\begin{figure}[ht]
    \centering
    \includegraphics[width=\columnwidth]{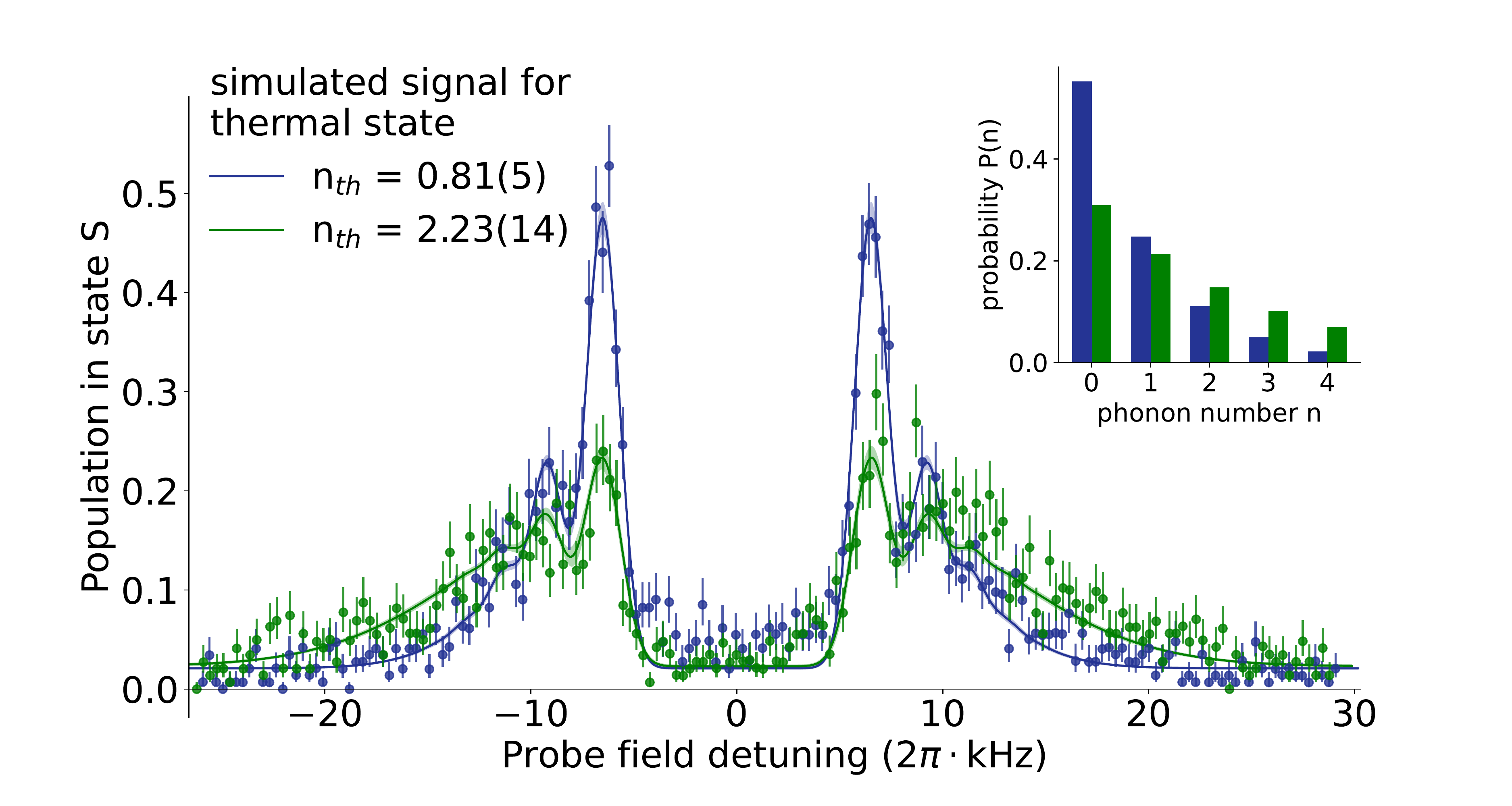}
\caption[width=\textwidth]{The Autler-Townes splitting shows multiple peaks for a thermal distribution of the ion motion. The peak position is defined by the phonon number and the amplitude by the population probability of this phonon number. By scanning the Autler-Townes spectrum the thermal distribution of the ion motion can be obtained.}
\label{fig_Thermal_dist}
\end{figure}

When the ion is prepared in a thermal state, the ion's motional state is a distribution of phonon number states. Each phonon number leads to a different splitting for the Autler-Townes doublet. Therefore, the thermal state manifests as a set of peaks at the respective phonon coupling strengths. The amplitudes of the different Autler-Townes peaks can be used to determine the probability of the respective phonon number to be occupied, and thus the phonon distribution of the thermal state can be characterized as shown in Fig.~\ref{fig_Thermal_dist}.

The thermal distribution in Fig.~\ref{fig_Thermal_dist} was created by shortening the cooling cycle before the Autler-Townes splitting scan, such  that the ion did not reach its motional ground state. With shorter cooling cycles, the ion is more likely to populate higher number states, leading to a higher average phonon number. Five Gaussian peaks  on each side of the Autler-Townes splitting are detected in Fig.~\ref{fig_Thermal_dist}, with the peak position defining the phonon number and the peak amplitude corresponding to the population. 

The resolving power of our technique is determined by the resonance linewidths and by the splitting between resonances.
The splitting between neighbouring resonances decreases as $\sim n^{-1/2}$, making the technique less powerful for larger $n$.
This drawback is common to other phonon measurement methods \cite{Meekhof1996, Leibfried1996, Shen2014, An2014, Um2016, Ding2017, Ohira2019, Meir2017}.
Larger splittings between resonances can be achieved by using a stronger coupling field, though this can also cause a higher background signal due to unwanted excitation of other levels. Furthermore, as the strength of the coupling field is increased, the AC-Stark shifts increase due to coupling to other levels and coupling field intensity fluctuations can cause broadening of the resonances \cite{Delone_1999}. The AC Stark shift can be compensated by adding an off-resonant field with the opposite detuning from the resonance \cite{Haffner2033}.

The spectral linewidth may be limited by the laser linewidth, magnetic field noise or Fourier broadening.
Longer probe times are required to reduce Fourier broadening, but this comes at the expense of increasing the sensitivity to anomalous heating which changes the ion's motional state \cite{Brownnutt2015}.

A further point to note is that the first-order description of the coupling strength scaling $\sim \sqrt{n+1}$ for BSB transitions and $\sim \sqrt{n}$ for RSB transitions breaks down when $\eta^2 (2n+1) \not \ll 1$, where $\eta$ is the Lamb-Dicke parameter.
In this regime the coupling strengths are best described using Bessel functions \cite{McCormick2019}.

However, an advantage of our method for resolving Fock states using the AC stark effect in comparison to the ones already measured for Rydberg atoms and superconducting circuits \cite{Brune1990,Guerlin2007, Schuster2007,Arrangoiz2019} might be the applicability in systems with weaker couplings. Schemes in superconducting circuits often require an ultra-strong coupling, due to the large detuning $\Delta$ with $g^2/\Delta\gg\kappa,\Gamma$, where $g=i\eta\Omega_P$, in order to dispersively resolve single photons. The decay rate is defined as $\Gamma$ and the field dissipation by $\kappa$. The strong coupling used in our scheme is achievable even in cavity QED systems \cite{Sames2018}, requiring only that $g^2\gg \kappa,\Gamma$.

\hfill

We introduced and demonstrated a measurement method for trapped ion motion in the number basis. It relies on the Autler-Townes effect where the splitting is dependent on the motional Fock state when coupling to a phonon number changing transition. We showed the expected scaling depending on BSB or RSB coupling and demonstrated that this method can be used to detect Fock states. The method can be repeatedly applied in a single experimental run until a positive result is achieved. Otherwise the method can enact a non-demolition measurement, and prepare trapped ions in phonon number states. A scan of the Autler-Townes spectrum allows one to determine the phonon number distribution of a thermal state.
Assuming that the excitation efficiency of the individual Autler-Townes doublet is high enough, one can also create a Fock state from a thermal distribution by using the Autler-Townes effect. The phonon number then defines the frequency of the laser, probing a well defined peak from the Autler-Townes separation, as done in Fig.~\ref{fig_fock_detect}(a) and (b). We have shown, that the measured scaling matches the theoretical model which scales as $\sqrt{n+1}$ for coupling of a BSB transition and $\sqrt{n}$ for coupling to a RSB transition.
For using the method, a system in which a quantum harmonic oscillator is coupled to a three level quantum system is needed. Hence, it can also be used in different systems such as superconducting circuits coupled to microwave cavities, or ions coupled to optical cavities, or atoms.

\hfill

 This work was supported by the Knut \& Alice Wallenberg Foundation (Photonic Quantum Information and the Wallenberg Centre for Quantum Technology [WACQT]), by the Swedish Research Council (Trapped Rydberg Ion Quantum Simulator, Grant No. 2020-00 381), and by the QuantERA ERA-NET Cofound in Quantum Technologies (ERyQSenS). The authors also thank the Joint Brazilian-Swedish Research Collaboration (CAPES-STINT), grant 88887.304806/2018-00 and BR2018-8054. R.B., M.H.O and C.J.V.-B thank the support from the National Council for Scientific and Technological Development (CNPq) grants 307077/2018-7, 311612/2021-0, 141247/2018-5, 409946/2018-4 and 313886/2020-2, and from the S\~ao Paulo Research Foundation (FAPESP) through Grants No.~2020/00725-9, 2019/13143-0, 2019/11999-5, and 2018/15554-5. This work has been supported by the French government, through the UCA$^\textrm{JEDI}$ Investments in the Future project managed by the National Research Agency (ANR) with the reference number ANR-15-IDEX-01.

\bibliography{main_arxiv}

\newpage
\section{Supplemental Material}
\subsection{Experimental setup}

The $\mathrm{^{88}Sr^+}$ ion is trapped in a linear Paul trap. During a single experimental cycle the ion is first Doppler cooled using the $5{}^2S_{1/2} \leftrightarrow 5{}^2P_{1/2}$ transition, then sideband cooling on the radial modes is applied to reach the motional ground state. Afterwards the ion is prepared in the state $\ket{S} \equiv 5{}^2S_{1/2} \:, m_J=-\frac{1}{2}$ by optical pumping.  Before the sequence of the Autler-Townes splitting, the ion is initialized in $\ket{D} \equiv 4{}^2D_{5/2} \:, m_J=-\frac{1}{2}$ via a $\pi$-pulse on the carrier transition.

\begin{figure}\centering
\includegraphics[width=\columnwidth]{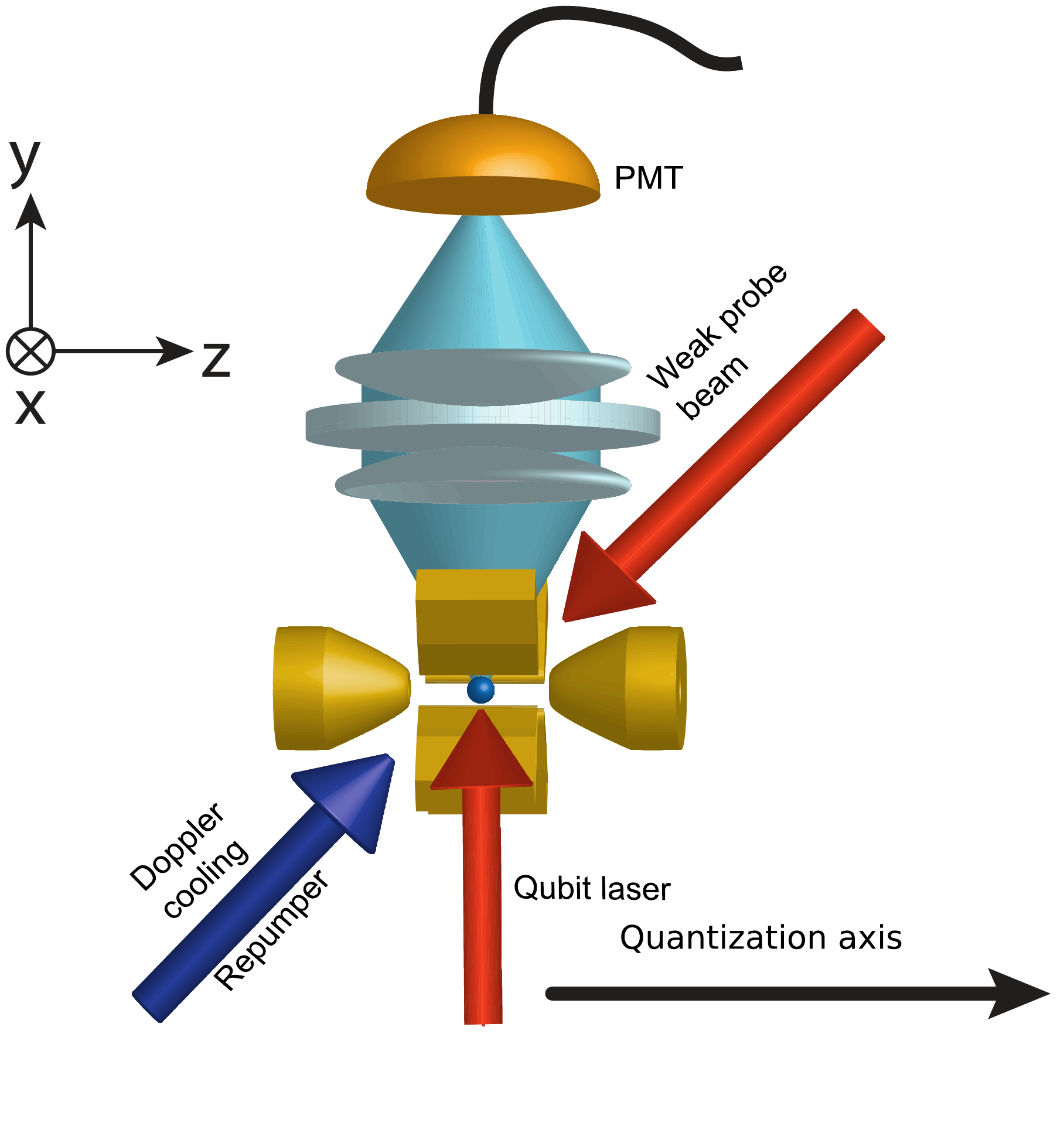}
\caption[width=\textwidth]{The ion is trapped using a linear Paul trap. The cooling, repump laser and the weak probe beam are under an $45^\circ$ angle to the trap axis. The remaining lasers for operations on the qubit transition are applied from the radial as well as then angled direction. Detection was done with a photomultiplier tube (PMT) mounted at the top of the experiment.}
\label{fig:sketch}
\end{figure}

For the weak probe, a laser beam at $45^\circ$ angle to the longitudinal trapping axis was used. The remaining pulses on the qubit transition were performed with a beam directed from the radial direction (at a $90^\circ$ to the trap axis). The fluorescence light is detected with a photomultiplier tube (PMT) mounted above the trap chamber. A sketch of the setup is shown in Fig.~\ref{fig:sketch}.
The fluorescence detection is done via the $5{}^2S_{1/2} \leftrightarrow 5{}^2P_{1/2}$ transition. Photons are therefore only detected if the ion is either in state $\ket{S}$ or in state $\ket{S'} \equiv 5{}^2S_{1/2} \:, m_J=\frac{1}{2}$. The ion decays during the detection sequence with a $\propto 6\%$ probability to the $4D_{3/2}$ state \cite{Zhang2018}. To minimize population losses in the experimental cycling transitions, a repump laser at 1092~nm, driving the ion population back to $5P_{1/2}$ state, is applied.

\subsection{Experimental parameter}

For scans involving the Autler-Townes splitting, a probe time of 700~$\mu$s was used. For the strong coupling pulses, in both cases, one of the two radial motional modes was used.

For the Autler-Townes scheme in which where the coupling was performed on a blue sideband (BSB) transition the Lamb-Dicke parameter was $\eta=0.0609$. The radial sideband which was used for the coupling was detuned from the carrier transition by $\Delta=2\pi\cdot1.3433$~MHz and the intitial coupling strength for the motional ground state was $\Omega_0=2\pi\cdot10.05(11)$~kHz.

The Lamb-Dicke parameter for the scans with red sideband (RSB) coupling was $\eta=0.0605$. The radial sideband was detuned by $-2\pi\cdot1.36$~MHz from the carrier transition and the coupling strength $\Omega_0=2\pi\cdot11.08(11)$~kHz in this case.

To create the thermal distribution of phonons in Fig.~4 in the main text, the length of the sideband pulse in the cooling sequence was reduced. The optimized cooling on the RSB transition was set to be at $2500$~$\mu$s. For the scan with $n_{th}=0.81(5)$ the cooling time was reduced to $500$~$\mu$s. The thermal distribution with $n_{th}=2.23(14)$ was produced by cooling only for $200$~$\mu$s.

\subsection{Fock state creation}
After the phonon number measurement sequence using the Autler-Townes splitting the ion ends up in a combination of the states $\ket{s,n}$ and $\ket{D,n+1}$. By applying a $\pi$ pulse on the $\ket{D',n+1} \rightarrow \ket{S,n}$ transition, followed by a $\pi/2$ pulse on the $\ket{D,n} \rightarrow \ket{S,n}$ transition, one can adapt the measurement scheme to create Fock states. Typically for the preparation of a particular phonon number state a post-selection step is performed after each BSB $\pi$ pulse. This method can be used with just one post-selection step after the excitation via the Autler-Townes splitting, thus making it more efficient for the preparation of phonon number states with larger $n$.

\subsection{Autler-Townes line shape}
Consider a three-level atomic system in a $V$ level configuration,
with a ground state $\vert S\rangle$ and excited states $\vert D\rangle$ and $\vert D'\rangle$. A probe field with frequency $\omega_{P}$ and Rabi frequency $\Omega_{P}$
couples the transition $\vert S\rangle\leftrightarrow\vert D\rangle$.
At the same time, a control field of Rabi frequency $\Omega_{0}$ and frequency
$\omega_{C}$ couples the transition $\vert S\rangle\leftrightarrow\vert D'\rangle$. The Hamiltonian describing this system can be written as $H=H_{0}+H_{int}$, with
\begin{equation}
H_{0}=\omega_{D'}\sigma_{D'D'}+\omega_{D}\sigma_{DD}+\nu a^{\dagger}a
\end{equation}
the term related to the atom's internal degrees of freedom
and the free energy of the motional mode. $\omega_{D}$ and $\omega_{D'}$ are the frequencies of the states $\vert D\rangle$ and $\vert D'\rangle$, respectively, and $\nu$ is the motional mode frequency. The second term of the total Hamiltonian, 
\begin{eqnarray}
H_{int} & = & \Omega_{P}\left(\sigma_{SD}+\sigma_{DS}\right)\left[e^{i\left(\omega_{P}t+\phi_{P}\right)}+e^{-i\left(\omega_{P}t+\phi_{P}\right)}\right]\nonumber\\
 &  & +\Omega_{0}\left(\sigma_{SD'}+\sigma_{D'S}\right)\left\{ e^{i\left[\eta\left(a^{\dagger}+a\right)-\omega_{C}t+\phi_{C}\right]}\right.\nonumber\\
 &  & \left.+e^{-i\left[\eta\left(a^{\dagger}+a\right)-\omega_{C}t+\phi_{C}\right]}\right\} 
\end{eqnarray}
is the one containing all the interacting terms, where $\sigma_{ii}$, with $i \in \{D,D'\}$ represents the population operators and $\sigma_{mn}=\vert m\rangle\langle n\vert$ the lowering and raising atomic operators, which promote transitions from state $\vert n\rangle$ to $\vert m\rangle$, for $m\neq n$ and $m,n \in \{S,D,D'\}$. $\eta$ is the Lamb-Dicke parameter, $\phi_{P}$ and $\phi_{C}$ are the phases of the probe and control field, respectively, and $a^{\dagger}$($a$) represents the creation (annihilation) operator acting on the Fock space. Here we have considered that the classical probe field does not couple the motional degrees of freedom of the atom. 

For convenience, we move to the interaction picture using the unitary transformation $U_{0}=e^{-iH_{0}t}$. Then, applying the Rotating Wave Approximation (RWA), in the Lamb-Dicke and low excitation regimes, i.e. $\eta\ll1$ and $\eta\sqrt{\langle (a^{\dagger}+a)^{2}\rangle}\ll1$, the Hamiltonian reads
\begin{eqnarray}
H_{I} & = & \Omega_{0}\sigma_{SD'}e^{i\left(\Delta_{C}t+\phi_{C}\right)}\left[1+i\eta\left(a^{\dagger}e^{i\nu t}+a e^{-i\nu t}\right)\right]\nonumber \\
 &  & +\Omega_{P}\sigma_{SD}e^{i\left(\Delta_{P}t-\phi_{P}\right)}+H.c.,
\end{eqnarray}
where $H.c.$ represents the Hermitian conjugate.

Choosing a particular phase and considering a resonant probe field, the Hamiltonian becomes
\begin{eqnarray}
H_{I} & = & \Omega_{0}\sigma_{SD'}e^{i\Delta_{C}t}\left[1+i\eta\left(a^{\dagger}e^{i\nu t}+ae^{-i\nu t}\right)\right]\nonumber \\
 &  & +\Omega_{P}\sigma_{SD}+H.c.\quad,\label{eq:H_lamb_dicke}
\end{eqnarray}
where we notice that there are three possible resonances for the coupling field: $\Delta_{C}=0$, which leads to a carrier transition where the number of excitations in bosonic mode is not affected by the control field, $\Delta_{C}=-\nu$, that couples to the first red sideband (RSB) transition, and $\Delta_{C}=\nu$, coupling the first blue sideband (BSB) transition. 

Next we obtain the lineshape for both the RSB and BSB transitions. It is important to stress that the linewidths predicted in this appendix are nothing more than a prediction for an ideal case. During the experiment, other sources of noise are present, such as fluctuations in laser intensity, pulse lengths, and motional heating. These noise sources cannot be neglected, and are generally predominant in the determination of the lineshapes.

\textit{Red-sideband (RSB)}: Considering $\Delta_{C}=-\nu$ , the
Hamiltonian in Eq.~\ref{eq:H_lamb_dicke} can be rewritten as 
\begin{equation}
H=\Omega_{P}\sigma_{SD}+ga^\dagger\sigma_{SD'}+H.c.,
\end{equation}
with $g=i\eta\Omega_C$. In the basis of
$n$ excitations in the bosonic mode $b=\left\{\vert S,n-1\rangle,\vert D,n-1\rangle,\vert D',n\rangle\right\} $
the three eigenvalues of this Hamiltonian are $E_{0}^{(n)}=0$ and
$E_{\pm}^{(n)}=\pm\sqrt{g^{2}n+\Omega_{P}^{2}}$, which means that the
system has a dark state with zero eigenenergy, and two symmetric eigenstates energetically separated by $2\sqrt{g^{2}n+\Omega_{P}^{2}}$. In a general form, these eigenstates can be written as
\begin{eqnarray}
\vert a_0^{(n)}\rangle & = & N_{n}^{a_0}\left[\vert D,n-1\rangle-\frac{\Omega_P}{g\sqrt{n}}\vert D',n\rangle\right],\nonumber \\
\vert\pm^{(n)}\rangle & = & N_{n}^{\pm}\left[\vert S,n-1\rangle \pm\left(\frac{\Omega _P}{\sqrt{g^2 n +\Omega_P^2 }}\vert D,n-1\rangle\right.\right.\nonumber \\
 &  & \left.\left.+\frac{g\sqrt{n}}{\sqrt{g^2 n +\Omega_P^2}}\vert D',n\rangle\right)\right],
\end{eqnarray}
with $N_{n}^{D}$ and $N_{n}^{\pm}$ being normalization factors. 

In the regime where the control field is much stronger than the probe
field, i.e., $g\sqrt{n}\gg\Omega_{P}$, these eigenenergies and eigenstates
become
\begin{eqnarray}
E_{0}^{(n)}=0 & : & \vert a_0^{(n)}\rangle=\vert D,n-1\rangle,\\
E_{\pm}^{(n)}=\pm g\sqrt{n} & : & \vert\pm^{(n)}\rangle=\frac{1}{\sqrt{2}}\left(\vert S,n-1\rangle\pm\vert D',n\rangle\right),
\end{eqnarray}
in accordance with the dressed state representation of the system. In this dressed state representation, the system is composed by $\vert D,n\rangle$ and the two symmetric states $\vert\pm^{(n)}\rangle$, with probe resonance at $\Delta_{P}=\pm g\sqrt{n}$. 

The lineshape width can be obtained, for a system with $n$ excitations
in the Fock space by calculating the total decay rate from one of
the dressed states $\vert\pm^{(n)}\rangle$ using Fermi's golden rule, with the respective collapse operators.
Concerning the spontaneous decay processes in the system, the total
decay rate is obtained considering the collapse operator $\sqrt{2\Gamma_{SD'}}\sigma_{SD'}$, resulting in 
\begin{eqnarray}
\Gamma_{\textrm{atom}} & = & \left|\langle S,n-1\vert\sqrt{2\Gamma_{SD'}}\sigma_{SD'}\vert +^{(n)}\rangle\right|^{2}\nonumber \\
& = & \Gamma_{SD'}.
\end{eqnarray}

For the contribution of the motional mode dissipation and heating in the lineshape width, we consider the collapse operators $\sqrt{2\kappa\left(n_{\mathrm{th}}^{\mathrm{env}}+1\right)}a$ and $\sqrt{2\kappa n_{\mathrm{th}}^{\mathrm{env}}}a^\dag$, with $n_\mathrm{th}$ being the mean number of environmental thermal phonons. So the total contribution is given by
\begin{eqnarray}
\kappa_{T} & = & \left|\langle S,n-2\vert\sqrt{2\kappa\left(n_{\mathrm{th}}^{\mathrm{env}}+1\right)}a\vert+^{(n)}\rangle\right|^{2}\nonumber \\
 &  & +\left|\langle D',n-1\vert\sqrt{2\kappa\left(n_{\mathrm{th}}^{\mathrm{env}}+1\right)}a\vert+^{(n)}\rangle\right|^{2}\nonumber \\
 &  & +\left|\langle S,n\vert\sqrt{2\kappa n_{\mathrm{th}}^{\mathrm{env}}}a^\dag\vert+^{(n)}\rangle\right|^{2}\nonumber \\
 &  & +\left|\langle D',n+1\vert\sqrt{2\kappa n_{\mathrm{th}}^{\mathrm{env}}}a^\dag\vert+^{(n)}\rangle\right|^{2}\nonumber \\
 & = & \kappa\left[2n\left(2n_{\mathrm{th}}^{\mathrm{env}}+1\right)-1\right].
\end{eqnarray}

Then, the spectral lineshape is fully characterized, with resonances
located at $\Delta_{P}=\pm g\sqrt{n}$ and with the $\text{FWHM}=\Gamma_{SD'}+\kappa\left[2n\left(2n_{\mathrm{th}}^{\mathrm{env}}+1\right)-1\right]$.

\textit{Blue sideband (BSB): }In the case when the control field is
resonant with the first blue sideband transition, i.e., $\Delta_{C}=\nu$,
the Hamiltonian from Eq.\ref{eq:H_lamb_dicke} becomes
\begin{equation}
H=\Omega_{P}\sigma_{DS}+ga\sigma_{SD'}+H.c..
\end{equation}

In the basis of $n$ excitations in the phonon mode $b=\left\{\vert S,n\rangle, \vert D,n\rangle,\vert D',n+1\rangle \right\} $,
we obtain, in
the regime where $g\sqrt{n+1}\gg\Omega_{P}$, the eigenvalues $E_{0}^{(n)}=0$ and $E_{\pm}^{(n)}=\pm g\sqrt{n+1}$, and  their respective eigenstates:
\begin{eqnarray}
\vert a_0^{(n)}\rangle&=&\vert D,n\rangle,\\
\vert\pm^{(n)}\rangle&=&\frac{1}{\sqrt{2}}\left(\vert S,n\rangle\pm\vert D',n+1\rangle\right),
\end{eqnarray}
where the only difference from the RSB transition is the effective
coupling strength being now $g\sqrt{n+1}$. Consequently, this will also alter the Autler-Townes splitting. 

The total spontaneous decay rate, considering the same collapse operator $\sqrt{2\Gamma_{SD'}}\sigma_{SD'}$, is given by
\begin{eqnarray}
\Gamma_{\textrm{atom}} & = & \left|\langle S,n+1\vert\sqrt{2\Gamma_{SD'}}\sigma_{SD'}\vert +^{(n)}\rangle\right|^{2}\nonumber \\
& = & \Gamma_{SD'},
\end{eqnarray}
while the contribution from the dissipation of the phonon mode, obtained with the collapse operators $\sqrt{2\kappa\left(n_{\mathrm{th}}^{\mathrm{env}}+1\right)}a$ and $\sqrt{2\kappa n_{\mathrm{th}}^{\mathrm{env}}}a^\dag$, is
\begin{eqnarray}
\kappa_{T} & = & \left|\langle S,n-1\vert\sqrt{2\kappa\left(n_{\mathrm{th}}^{\mathrm{env}}+1\right)}a\vert+^{(n)}\rangle\right|^{2}\nonumber \\
 &  & +\left|\langle D',n\vert\sqrt{2\kappa\left(n_{\mathrm{th}}^{\mathrm{env}}+1\right)}a\vert+^{(n)}\rangle\right|^{2}\nonumber \\
 &  & +\left|\langle S,n+1\vert\sqrt{2\kappa n_{\mathrm{th}}^{\mathrm{env}}}a^\dag\vert+^{(n)}\rangle\right|^{2}\nonumber \\
 &  & +\left|\langle D',n+2\vert\sqrt{2\kappa n_{\mathrm{th}}^{\mathrm{env}}}a^\dag\vert+^{(n)}\rangle\right|^{2}\nonumber \\
 & = & \kappa\left[2n\left(2n_{\mathrm{th}}^{\mathrm{env}}+1\right)+4n_{\mathrm{th}}^{\mathrm{env}}+1\right].
\end{eqnarray}

This means that in the case of a BSB transition, the spectral lineshape
is characterized by resonances located at $\Delta_{P}=\pm g\sqrt{n+1}$
and a $\text{FWHM}=\Gamma_{SD'}+\kappa\left[2n\left(2n_{\mathrm{th}}^{\mathrm{env}}+1\right)+4n_{\mathrm{th}}^{\mathrm{env}}+1\right]$.

\end{document}